# Origin of band gaps in 3d perovskite oxides

Julien Varignon[1], Manuel Bibes[1] and Alex Zunger[2]

[1]*Unité Mixte de Physique, CNRS, Thales, Université Paris Sud, Université Paris-Saclay, 91767, France*

[2]*University of Colorado Boulder Colorado 80309, Boulder, CO, USA*

*Abstract*

With their broad range of magnetic, electronic and structural properties, transition metal perovskite oxides $ABO_3$ have long served as a platform for testing condensed matter theories. In particular, their insulating character – found in most compounds – is often ascribed to dynamical electronic correlations through the celebrated Mott-Hubbard mechanism where gaping arises from a uniform, symmetry-preserving electron repulsion mechanism. However, structural distortions are ubiquitous in perovskites and their relevance with respect to dynamical correlations in producing this rich array of properties remains an open question. Here, we address the origin of band gap opening in the whole family of *3d* perovskite oxides. We show that a single-determinant mean-field approach such as density functional theory (DFT) successfully describes the structural, magnetic and electronic properties of the whole series, at low and high temperatures. We find that insulation occurs via energy-lowering crystal symmetry reduction (octahedral rotations, Jahn-Teller and bond disproportionation effects), as well as intrinsic electronic instabilities, all lifting orbital degeneracies. Our work therefore suggests that whereas $ABO_3$ oxides may be complicated, they are not necessarily strongly correlated. It also opens the way towards systematic investigations of doping and defect physics in perovskites, essential for the full realization of oxide-based electronics.

## I. INTRODUCTION

The striking range of experimentally observed low temperature (LT) and high temperature (HT) magnetic, electronic, and structural properties across the *3d* ABO$_3$ perovskites family have held the condensed matter physics community in constant fascination for many years [1–4]. Regarding **magnetism,** the LT phases are generally spin-ordered, either ferromagnetic (FM) (*e.g.* YTiO$_3$ [5]) or antiferromagnetic (AFM) (*e.g.* LaVO$_3$ [6], CaMnO$_3$ [7], LaMnO$_3$ [8], CaFeO$_3$ [9], LaFeO$_3$ [8] or YNiO$_3$ [10]), whereas the HT phases exhibit spin-disordered paramagnetism (PM). Regarding the electronic **metal vs insulator band gap characteristics**, these compounds show three modalities: (i) most LT and HT phases are insulating, except (ii) CaVO$_3$ and SrVO$_3$ that are said to be PM metals at all temperatures [11], whereas (iii) CaFeO$_3$ [9] or YNiO$_3$ [10,12] display both PM metal and PM insulator HT phases. Regarding **structural** aspects, these perovskites show a range of structure-types (cubic, monoclinic, orthorhombic) as well as structural distortions within such space groups, (including octahedral rotations, anti-polar displacements), as well as Jahn-Teller (LaVO$_3$, LaMnO$_3$) or breathing distortions (CaFeO$_3$ and YNiO$_3$) types (see Supplementary S1 for sketches of the distortions). The HT PM phases usually inherit the LT structure, and sometimes they can also transform to their own, distinct structure-type (as in LaVO$_3$, CaFeO$_3$, YNiO$_3$).

The enormous number of publications in this field typically focus on one or just few selected ABO$_3$ compound(s) and one or two of the effects noted above. This makes it difficult to assess the key question: *what is the minimal physical description needed to capture the basic magnetic, transport and structural ground state properties, and can one define a single, overarching theoretical framework that works essentially across the boar*d?

**The question of the origin of Mott band gaps:** Perhaps the most celebrated issue in this regard, raised by Peierls and Mott [3,4], concerns the way in which a band gap can form in the presence of an odd number of electrons, when the Fermi energy E$_F$ would intersect what appears to be a partially occupied, gapless band [3]. The classic Mott-Hubbard 'strongly correlated' symmetry-conserving view has formed a central paradigm for teaching and explaining the known phenomenology [13–15]. This view was motivated by the fact that in the extreme ionic limit (where the *3d* ion carries all of the active electrons, while O$^{2-}$ is a rigid, closed-shell), one would have guessed that the single-particle levels near $E_F$ would be degenerate with partial occupancy (such as the triply degenerate $t_{2g}$ occupied by just 2 electrons, or the doubly degenerate $e_g$ occupied by just a single electron). The Mott-Hubbard *mechanism* for gap formation and magnetism in such *d*-electron oxides (Fig.1.a) thus envisions electrons moving across the lattice, forming atomic-like states on certain *3d* atomic sites with doubly occupied *d* orbitals ('upper Hubbard band') and empty *d* orbitals on other sites ('lower Hubbard band'). In this picture, the band gap of the AFM and PM phases of these oxides therefore emerges due to this correlation-induced *electron-electron repulsion* (Fig.1.b top panel), without spatial symmetry breaking (such as structural distortions, or magnetic order that would lower the symmetry of the degenerate states). Symmetry can break later, in this view, as a non-essential step. From this point of view, the minimal theoretical description needed would entail multi-determinant dynamically correlated methodologies capable of treating open shell degenerate configurations. It is noteworthy that these methods focused on metal/nonmetal band gap aspects but have rarely predicted the aforementioned *observed* structural distortions or examined their relationship to gaping.

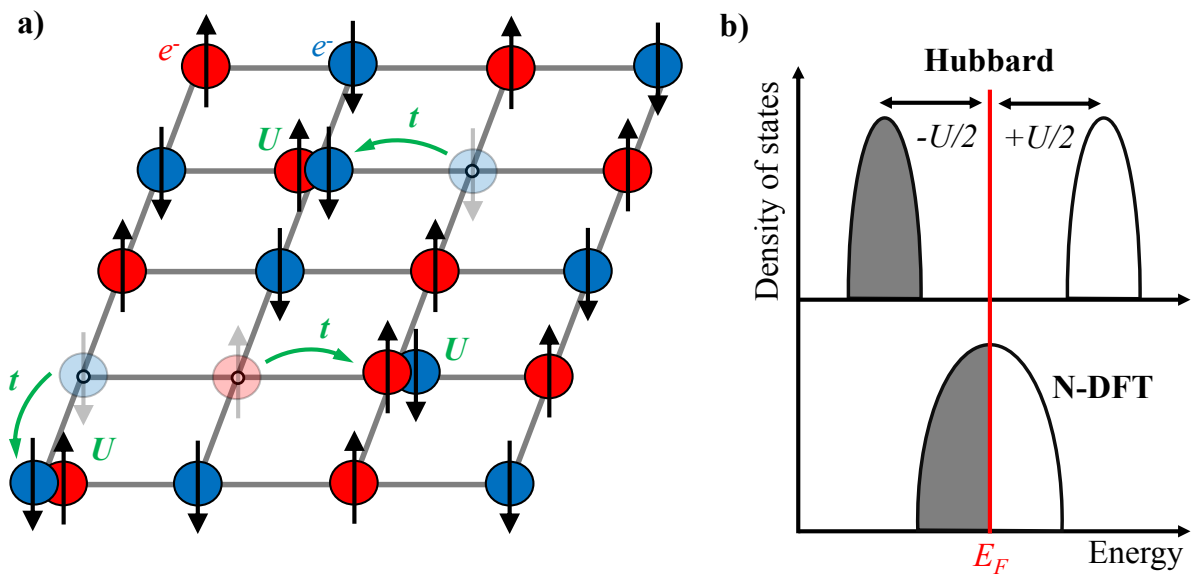

*Figure 1:* a) Mott-Hubbard mechanism with one *d* electron per site having either a spin up or a spin down. Assuming no symmetry breaking, electrons are hoping on the lattice with a probability *t* and produce doubly occupied (upper Hubbard band) and empty (lower Hubbard band) orbitals on the sites. Double occupancy causes energy penalty U. (b). The naïve Density functional theory (N-DFT, lower panel) assuming also no symmetry breaking would have instead an ungaped, metallic band.

***The failure of naïve DFT models does not disqualify DFT***: A natural adoption of this symmetry-conserving principle within band theory for PM phases is to assume a structure where all *3d* ions are symmetry-equivalent, having identical local environments (a *monomorphous representation*) while, (as expected for the macroscopic PM phase), the total magnetic moment is zero. But such a non-magnetic description forces zero moment *on an atom-by-atom basis*, and so produces zero gap in a band theoretic description for systems with odd number of electrons (Fig.1b bottom part), in stark contrast with experiment for most PM phases (Fig.2). Unfortunately, the DFT total energy of this non-magnetic (NM) approximation is far higher than what is obtained with DFT in a proper spin-polarized PM ground state (by 452, 1167, 935, 4313, and 1925 meV/f.u for $YTiO_3$, $LaVO_3$, $CaMnO_3$, $LaMnO_3$, and $CaFeO_3$, respectively, see Fig.2 and supplementary S2 for further details). Even though such naïve DFT approximations correspond to extremely high-energy solutions, they were often used in the literature to suggest that band theory fails to explain gaping of Mott insulators, the latter being argued to require instead an explicitly correlated approach [16–20]. Since this argument is evidently untested in the context of non-naïve DFT, one must look further for what is the minimal theoretical framework that does work.

***DFT without U can also explain Mott gaps:*** A number of calculations have used the 'DFT+U' approach, where DFT is amended by an on-site potential that removes part of the spurious self-interaction error and thereby creates a distinction between occupied and empty states–producing at times gaped states [21–29]. This success has propagated the view that it is the interelectronic repulsion akin to the Hubbard Hamiltonian that produces gaping in DFT+U. However, the role of the on-site potential U in DFT+U, where U is a one body on-site potential shifting the *d* orbital to deeper energies, is distinct from the Hubbard Hamiltonian, where it truly represents interelectronic repulsion.

In order to settle the issue if the U parameter is crucial for the gapping in the transition metal oxide perovskites, we performed independent calculations for all experimentally observed $ABO_3$ PM insulators considered in this paper (Fig.2) using the meta GGA SCAN functional [30] (satisfying 17 important boundary conditions for XC) *without any U.* We find (details in Supplementary S3) in all cases that compounds found insulators within DFT+U are also gapped using U=0 eV meta GGA and that there are strong similarities between the density of states (DOS) of DFT+U and SCAN-no-U. This further substantiates the fact that U parameter is not really mandatory to produce insulation in oxide perovskite. *We conclude that the question of what is the minimal theory that describes the basic ground state phenomenology across the $ABO_3$ series—symmetry broken or symmetry conserving; or statically correlated (as in DFT), or requiring dynamic correlation-- is still unsettled.*

In the present paper we show that a single-determinant mean-field approach such as density functional theory (DFT) – that makes no use of the extreme ionic view of the bonding and allows the system to select its own *total energy lowering* structures, forms of magnetism and the ensuing formation of closed orbital sub-shells – successfully describes the basic properties across the *3d* perovskite series of materials, including both LT and HT phases (Fig.2). Instead of implying a universal role for dynamical correlations, gapping is often achieved through symmetry breaking via energy-lowering reduced crystal symmetries (octahedral rotations, Jahn-Teller and bond disproportionation effects) and electronic instabilities, all lifting orbital degeneracies. This is a significant result because it suggests that (a) whereas $ABO_3$ oxides may be (structurally and magnetically) complicated, they are not necessarily strongly correlated, and (b) a rather simple tool such as DFT (requiring a low computational effort with respect to heavier machineries treating dynamical correlations) offers a single platform to study reliably and with sufficient precision not only band gap formation, structure and magnetism in $ABO_3$, but also – in the future – doping, defect physics and interface effects.

## II. Approach

The dual input to this (or other) theories is: (a) a framework for interelectronic interaction (here, DFT) and (b) a description of the structural and electronic degrees of freedom that could be optimized. Previously, technical shortcomings in (b) within a naïve DFT approach were often attributed to theoretical failures in describing the underlying interelectronic interactions (a).

(a) **The DFT format used:** Since we are interested in determining what are the minimum theoretical ingredients needed to explain the ground state properties observed in the $ABO_3$ series, we deliberately adopt for (a) a mean-field, single determinant, Bloch periodic band structure approach, with an electron-gas based description of exchange and correlation (XC). Building on the central fact that DFT is an exact formal theory for the ground-state properties for the *exact* exchange-correlation energy functional, there is no reason in principle why the properties noted above couldn't be captured by the ultimate DFT. The reason for the limitations of currently used DFT is that the description of exchange and correlation (XC) is imperfect. The XC functional represents a hierarchy of approximations [30,31] ("rungs"). The first three rungs correspond to an XC potential depending on the local density (LDA) and its first (GGA) and second derivative (meta GGA), being local or semi-local functionals of the noninteracting density matrix, with no distinction between occupied and unoccupied levels. The first rung that distinguishes occupied from unoccupied states is rung 4 being a nonlocal functional of the non-interacting density matrix. While ideally this would be a fully self-interaction corrected functional, this functional does not exist as yet, so we use a simple approximation to it in form of the DFT+U method. Further details on the DFT calculations and on the choice of U parameters are provided in Supplementary S4.

(b) ***The structural and electronic degrees of freedom that could be optimized***: In order to provide a real test to the ability of the DFT XC functional to describe the basic trio of properties—spin arrangement, gaping and structure—it is obviously necessary to provide in (b) sufficient electronic and structural generality and flexibility *in the way the system is represented* so that symmetry breaking events requiring such flexibility could be captured insofar as they lower the total energy. Here, instead of restricting the unit cell representation of PM (and for AFM) magnetic structures to a single, primitive cell, we allow repeated multiples of the underlying cubic, orthorhombic or monoclinic ("super") cells, subject to the constraint that the *overall structure* represents the macroscopic crystal and spin structures. We use the Special Quasi Random (SQS) [32] construct that selects supercells of given crystalline space group symmetry such that the occupation of sites by spins follows a random pair and multi body correlation functions (appropriate to the HT limit) with a total moment of zero (see Supplementary S5 for details on SQS). This polymorphous representation [21] provides *an opportunity* to break spatial symmetry, should the total energy be lowered in doing so.

The symmetry breaking channels explored here include:

(a) **Allowing *different local environments*** for the various chemically identical *3d* atoms in the lattice. Specifically, a different number of spin-up vs spin-down sites can exist around each *3d* site subject to the SQS constraint that all pair interactions are purely random (*i.e.* no short-range order) and the total spin is zero.

*(b) **Occupation number fluctuations*** whereby atomic sites with partial occupation of initially degenerate levels can have different assignments of the electrons to the degenerate partners [such as (1,0,1) for 2 electrons in the 3 partners of $t_{2g}$, rather than using fractional and equal occupation such as (2/3,2/3,2/3)] [33].

(c) **Displacement fluctuations** (*i.e.* atomic relaxation requiring cells larger that the primitive cell), including local Jahn-Teller distortions as well as octahedral mode deformations such as breathing, tilting, rotation, and anti-polar displacements described in details in Supplementary S1.

### III. Results

#### a. Predictions for low temperature magnetically ordered AFM/FM phases

Figure 2 summarizes our results for the lowest energy phase considering spin-ordered phases (energy differences between all tested magnetic configurations and initial symmetries are given in Table SI4 of Supplementary S6). Consistent with experiments and previous DFT theoretical literature [22–28,34], for all explored compounds we find (i) the correct LT crystal structure – orthorhombic for $YTiO_3$, $CaVO_3$, $CaMnO_3$, $LaMnO_3$, $LaFeO_3$; monoclinic for $LaVO_3$, $CaFeO_3$ and $YNiO_3$; cubic for $SrVO_3$; (ii) the correct low T spin ordered phase including AFM (for $YTiO_3$, $LaVO_3$, $CaMnO_3$, $LaMnO_3$, $LaFeO_3$, $YNiO_3$) or FM ( for $YTiO_3$) (except for $CaFeO_3$ that exhibits an incommensurate antiferromagnetic spin spiral order at low temperature [9] not included in our modeling); (iii) all compounds adopting a spin-ordered ground state ($YTiO_3$, $LaVO_3$, $CaMnO_3$, $LaMnO_3$, $CaFeO_3$, $LaFeO_3$ and $YNiO_3$) are predicted insulating. Furthermore, (iv) the key *cell-internal* lattice distortions ($O_6$ group rotations, Jahn-Teller distortions, bond disproportionation) observed experimentally, are reproduced by theory with mode amplitudes in excellent agreement with experiments (see Table SI5 in Supplementary S7).

#### b. Predictions for High T Paramagnetic phases

***Predicted structure-types and sublattice distortions***. Using the polymorphous representation, the DFT+SQS reproduces the experimentally observed structure for PM phases ($YTiO_3$, $CaVO_3$, $SrVO_3$, $LaVO_3$, $CaMnO_3$, $LaMnO_3$, $CaFeO_3$, $LaFeO_3$, $YNiO_3$) as summarized in Fig.2. Interestingly, the relaxed HT structures share key similarities with the LT phases ($O_6$ rotations, bond disproportionation) as inferred by our symmetry-adapted mode analysis (presented in Table SI5 of Supplementary S7), *i.e.* the LT phases generally inherits the properties of the PM phases. Only LaVO$_3$ exhibits a PM phase that undergoes an alternative Jahn-Teller motion pattern with respect to the LT phase, yielding an orthorhombic symmetry instead of a monoclinic symmetry (Table SI5 and Fig.SI1 in Supplementary S7 and S1, respectively).

***Predicted metal vs insulate characteristics:*** The present approach correctly reproduces the experimentally observed insulating or metallic behaviors for all materials (Fig.2). Surprisingly, it also correctly reproduces the two experimentally observed paramagnetic phases of CaFeO$_3$ and YNiO$_3$, with an insulating phase slightly more stable than the metallic phase (Fig.2 and Table SI4 in Supplementary S6). Finally, we note in passing that ABO$_3$ with early 3*d* element B=Ti, V show band edges that are *d*–like (Upper Hubbard and lower Hubbard bands, candidates for Mott behavior (*c.f* Fig.1.b top), whereas those with later *3d* element (B=Mn, Fe, Ni) display oxygen-like band edges (see Figs.3, 4, 5 and 6), generally not necessarily Mott like.

|  |  | CaMnO$_3$ | LaFeO$_3$ | YTiO$_3$ | CaVO$_3$ | SrVO$_3$ | LaMnO$_3$ | LaVO$_3$ | CaFeO$_3$ | YNiO$_3$ |
|---|---|---|---|---|---|---|---|---|---|---|
|  |  | $t_{2g}^3 e_g^0$ | $t_{2g}^3 e_g^2$ | $t_{2g}^1 e_g^0$ | $t_{2g}^1 e_g^0$ | $t_{2g}^1 e_g^0$ | $t_{2g}^3 e_g^1$ | $t_{2g}^2 e_g^0$ | $t_{2g}^3 e_g^1$ | $t_{2g}^6 e_g^1$ |
|  | t factor | 0.98 | 0.94 | 0.87 | 0.96 | 1.02 | 0.94 | 0.89 | 0.89 | 0.92 |
|  | Gapping Mechanism | Octahedral crystal field splitting | | Symmetry lowering | | | | Symmetry lowering + Jahn-Teller effect | Disproportionation effects | |
| LT phase | Structure | Pbnm ✓ | Pbnm ✓ | Pbnm ✓ | Pbnm ✓ | Pm-3m ✓ | Pbnm ✓ | P2$_1$/b ✓ | P2$_1$/n ✓ | P2$_1$/n ✓ |
|  | Magnetism | AFM-G ✓ | AFM-G ✓ | FM ✓ | PM ✓ | PM ✓ | AFM-A ✓ | AFM-C ✓ | Spiral ✗ (FM) | AFM-S ✓ |
|  | Met-Ins | Ins. ✓ | Ins. ✓ | Ins. ✓ | Metal ✓ | Metal ✓ | Ins. ✓ | Ins. ✓ | Ins. ✓ | Ins. ✓ |
|  | E$_g$ (eV) / m (μ$_B$) | 0.69 / 2.49 | 2.31 / 4.12 | 1.11 / 0.94 | 0 / 0.79 | 0 / 0.78 | 1.86 / 3.81 | 1.73 / 1.91 | 0.11 / 3.85-3.42 | 0.50 / 1.19-0.00 |
|  | ΔE$_{LT-HT}$ (meV/f.u) | -23 | -118 | -10 | 0 | 0 | -4 | -10 | -52 | -4 |
| HT PM phase | Structure | Pbnm ✓ (NM: ✓) | Pbnm ✓ (NM: ✓) | Pbnm ✓ (NM: ✓) | Pbnm ✓ (NM: ✓) | Pm-3m ✓ (NM: ✓) | Pbnm ✓ (NM: ✓) | Pbnm ✓ (NM: ✗) | 1: P2$_1$/n ✓ / 2: Pbnm ✓ (NM 1: ✗) | 1: P2$_1$/n ✓ / 2: Pbnm ✓ (NM 1: ✗) |
|  | Met-Ins | Ins. ✓ (NM: ✗) | Ins. ✓ (NM: ✗) | Ins. ✓ (NM: ✗) | Metal ✓ (NM: ✓) | Metal ✓ (NM: ✓) | Ins. ✓ (NM: ✗) | Ins. ✓ (NM: ✓) | 1: Ins. ✓ / 2: Metal ✓ (NM 1: ✗) | 1: Ins. ✓ / 2: Metal ✓ (NM 1: ✗) |
|  | ΔE$_{PM-NM}$ (meV/f.u) | -935 | -1997 | -442 | -53 | -63 | -4313 | -1167 | -1925 | -130 |
|  | E$_g$ (eV) / m (μ$_B$) | 0.64 / 2.52 | 1.31 / 4.19 | 1.26 / 0.90 | 0 / 0.79 | 0 / 0.78 | 1.74 / 3.75 | 1.60 / 1.91 | 1: 0.06 / 3.94-3.07  2: 0 / 3.59 | 1: 0.50 / 1.26-0.16  2: 0 / 0.77 |

***Figure 2:*** Summary of the properties (electronic configurations, space group symmetry, metal vs. insulator behavior, spin order, Goldschmidt tolerance factor, mode of gapping, energy differences between the different magnetic solutions, band gap and magnetic moments) of oxide perovskites with unpaired electronic d shell configurations tested in our simulations. Green (red) tick marks refer to DFT success (failure) to reproduce the experimental

observations available in literature. For the high temperature phase, the results of the Non-Magnetic approximation to PM are provided in parenthesis.

## IV. Gapping mechanisms

Having now established that the methodology used correctly captures the basic physical properties observed experimentally for the different compounds presented in Fig.2, we next study the factors causing gaping, providing a *classification of gaping mechanisms for the different ABO$_3$ perovskites.*

The methodology used to identify the leading gaping mechanism is to start from an *assumed* high symmetry cubic perovskite phase (*Pm-3m* symmetry), and slightly 'nudge' (in the linear response sense) this structure with respect to potential symmetry breaking modes discussed in Sec. II, looking for energy lowering and gap formation. The perturbing modes are (a) allowing different local environments, (b) occupation number fluctuation, and (c) displacement fluctuations including local Jahn-Teller distortions as well as octahedral mode deformations such as breathing, tilting, rotation, and anti-polar displacements. Once the leading symmetry breaking mechanism is found (*wrt* cubic), we proceed to perform the complete gaping and magnetism calculation using the actual crystal structure (orthorhombic, cubic and monoclinic) looking for minimum energy. We group the ABO$_3$ compounds into 4 categories in terms of the gaping mechanisms.

### a. Mechanism I: Compounds forming closed subshells (no electronic degeneracies) by octahedral breaking of atomic symmetry: CaMnO$_3$ and LaFeO$_3$

The *free ions* Mn$^{4+}$ ($d^3$) and Fe$^{3+}$ ($d^5$) with odd number of electrons would simplistically lead to partially filled degenerate states and hence candidate for ungaped "metallic bands". But the solid-state octahedral symmetry breaks the continuous rotation atomic *d* symmetry: Starting from an assumed cubic phase, CaMnO$_3$ and LaFeO$_3$ are already found to be insulators with band gaps of 0.10 eV (CaMnO$_3$, $t_{2g}^3$) and 0.85 eV (LaFeO$_3$, $t_{2g}^3 e_g^2$) and sizable magnetic moments of 2.46 $\mu_B$ and 4.20 $\mu_B$, respectively (Figs.3.a and b). Consequently, in this group the band gap originates from lowering the atomic symmetry by the octahedral field and Hund's rule, driving half-filling of the degenerate partners and an ensuing gap already at the cubic level. Because of the relative A-to-B atomic size mismatch in $ABO_3$ (as reflected by the 1926 Goldschmidt tolerance factor $t$ [35] reported in Fig.2, being less than 1), these compounds do not prefer the ideal cubic structure and could distort by $O_6$ rotations, as indeed obtained by DFT energy minimization. This additional symmetry lowering produces energy gains of 315 and 378 meV/f.u in $CaMnO_3$ and $LaFeO_3$, respectively, further increasing their band gaps to 0.64 and 1.31 eV without altering the B cations magnetic moment (Fig.2).

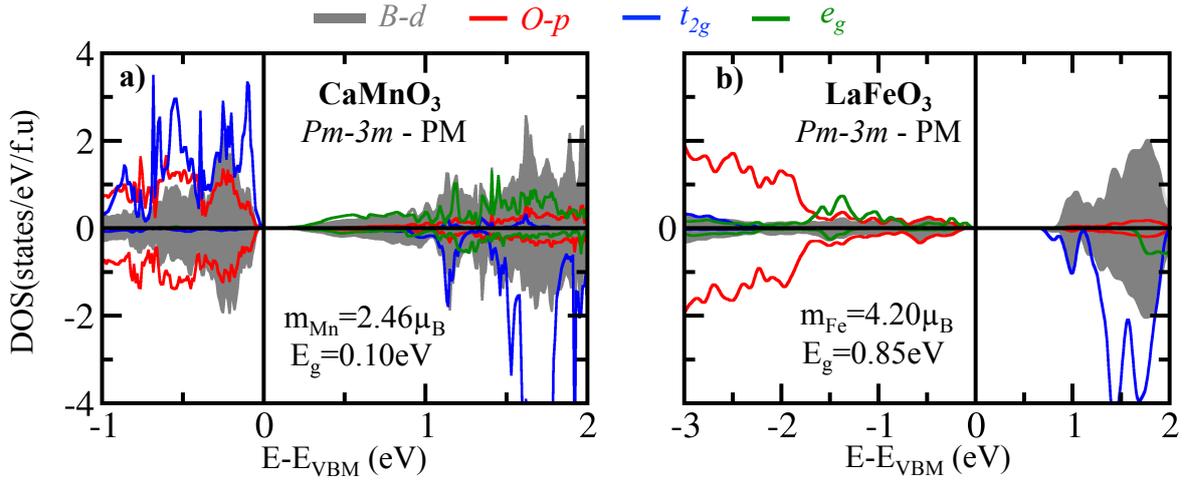

*Figure 3*: Averaged projected density of states on B d levels (grey) and O p levels (red) in $CaMnO_3$ (upper panel) and $LaFeO_3$ (lower panel) in a hypothetical cubic phase within the PM order. Projected density of states on $t_{2g}$ (blue) and $e_g$ (green) levels for a specific B cation within the supercell are also reported.

   b. *Mechanism II: Compounds lifting electronic degeneracies due to octahedral rotations: (YTiO₃ and LaMnO₃) and those retaining electronic degeneracies due to small or absent octahedral rotations (CaVO₃ and SrVO₃)*

Unlike $CaMnO_3$ and $LaFeO_3$, the cubic symmetry alone does not create half filling in $CaVO_3$, $SrVO_3$, $YTiO_3$ ($d^1$) and $LaMnO_3$ ($d^4$) that continue to exhibit electronic degeneracies of either $t_{2g}$ or $e_g$ levels. $CaVO_3$ and $SrVO_3$ are metals whereas $YTiO_3$ and $LaMnO_3$ are insulators in both PM and LT phases. These can be classified in terms of the strength of the Octahedral tilting effects:

**$YTiO_3$ and $LaMnO_3$ are gaped insulators due to octahedral rotation in the absence of electronic instability**: We first probed the $d^1$ orbital degeneracy of $Ti^{3+}$ in $YTiO_3$ and of $d^4$ of $Mn^{3+}$ in $LaMnO_3$ by nudging the occupancy to a specific degenerate partner *within the cubic* phase (e.g. occupation of the $t_{2g}$ triplet by (1,0,0) rather than (1/3,1/3,1/3)). We find that the imposed orbital broken symmetry (OBS) is unstable and decays back to a metallic solution with equal occupations of degenerate partners. Thus, in the ideal cubic symmetry these compounds do not develop any electronic instability that would break the $d$ orbital degeneracy thereby leading to gap formation.

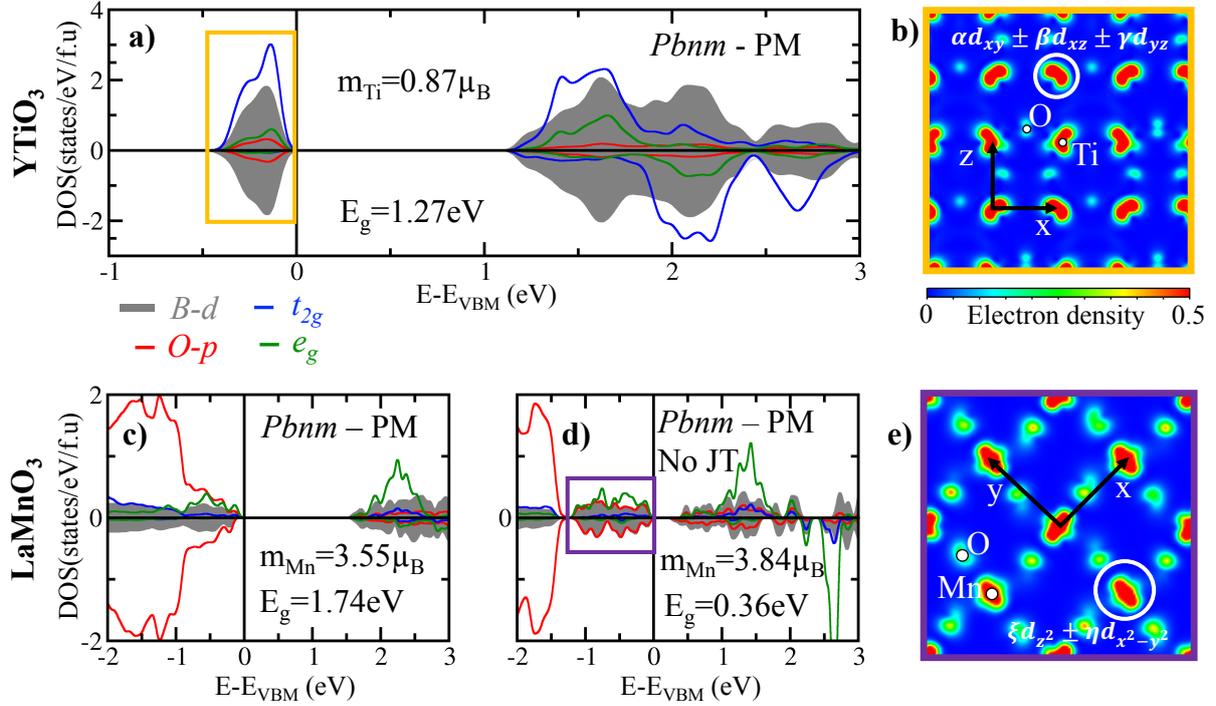

*Figure 4*: (a,c,d) Averaged projected density of states on B d levels (grey) and O p levels (red) in YTiO$_3$ (B=Ti, upper panel) and LaMnO$_3$ (B=Mn, lower panel) in the PM phase. Projected density of states on $t_{2g}$ (blue) and $e_g$ (green) levels for a specific B cation within the supercell are also reported. (b,e) Partial charge density maps in the (ab)-plane of levels located near the Fermi level reported in a and d.

Permitting next non-cubic symmetry breaking reveals the gapping mechanism here. Due to a large A-to-B cation size mismatch in ABO$_3$, reflected by their tolerance factor (Fig.2), YTiO$_3$ and LaMnO$_3$ are expected to be unstable in their cubic structures and to develop O$_6$ rotations, lowering the symmetry from cubic *Pm-3m* to orthorhombic *Pbnm,* as indeed found in DFT energy minimization. This results in large total energy lowering (ΔE=-1898 and -632 meV/f.u in YTiO$_3$ and LaMnO$_3$, respectively) and in insulating phases with a band gap created between *d* levels (Figs.4.a and c). Due to the symmetry lowering, the point group symmetry is reduced from O$_h$ to D$_{2h}$ and a new basis of *d* orbitals is produced locally on each transition metal sites. In YTiO$_3$, it results in a split-off *d* band below the Fermi level (Fig.4.a) and the Ti *d* electron is localized in an orbital corresponding to a linear combination of the "cubic-$t_{2g}$" levels (Fig.4.b) of the form $\alpha d_{xy} + \beta d_{xz} + \gamma d_{yz}$ ($\alpha^2 + \beta^2 + \gamma^2$=1), whose coefficients $\alpha$, $\beta$ and $\gamma$ are triggered by combinations of octahedra rotations and A cations anti-polar motions [24] (no Jahn-Teller (JT) motions are observed in the structure on the basis of our symmetry adapted modes presented in Table SI5 in Supplementary S7. This results in "orbital ordering" that is clearly a DFT band structure effect triggered by lowering the site symmetry, not a correlation effect. The characteristic shape of the orbital (reflected in the charge-density map of Fig.4.b) then drives the FM interactions at low temperature [24].

A similar mechanism applies for LaMnO$_3$, although the material displays a large in-phase JT distortion (Table SI5). This produces a gap located between $e_g$ levels (Fig.4.c). To settle the importance of the additional JT motion *wrt* YTiO$_3$, we calculate the very same *Pbnm* phase but remove the JT mode. We observe that the material is already an insulator and the averaged Mn$^{3+}$ magnetic moments are mostly unaltered (Fig.4.d). Therefore, O$_6$ rotations are sufficiently large to split the new basis of $e_g$ levels and the Mn$^{3+}$ electrons are localized in a cigar shape orbital pointing either along x or y directions on Mn neighboring sites in the (*ab*)-plane (Fig.4.e). These cigar shape orbitals correspond to linear combinations of the "cubic $e_g$" levels of the form $\xi d_{z^2} + \eta d_{x^2-y^2}$ ($\xi^2 + \eta^2$=1, $|\xi| \approx |\eta|$). The specific orbital pattern stabilizes an AFM-A order at low temperature, without changing substantially the band gap and the averaged magnetic moments (Fig.2). Finally, we note that although any electronic instability was identified, the in-phase Jahn-Teller motion likely originates from lattice mode couplings with oxygen rotations [22,36,37] (see Supplementary S8).

**CaVO$_3$ and SrVO$_3$ are metals because of insufficiently large O$_6$ rotations:** Because of their closer to 1 tolerance factor (Fig.2), CaVO$_3$ and SrVO$_3$ are not developing large O$_6$ rotations (TableSI5), as found in DFT energy minimization, and thus the orbitals are not sufficiently split to render an insulating phase. Note that the metallic state is not due to some strongly correlated effect but rather due to a trivial, semi classical factor captured by the 1926 Goldschmidt tolerance factor [35]. The mechanism was validated by artificially imposing the YTiO$_3$-type O$_6$ rotations onto CaVO$_3$, finding in our variational self-consistency, an insulator with a band gap of 0.14 eV. In agreement with this result, Pavarini *et al* noted in their Dynamical Mean Field Theory (DMFT) simulations [19,38,39] on YTiO$_3$ and LaMnO$_3$ that orthorhombic distortions (rotations, A-cations motions) can produce a sufficiently large crystal field splitting localizing electrons, irrespective of Jahn-Teller distortions.

### c. Mechanism III: Compounds with 2 electrons in $t_{2g}$ levels: the case of insulating LaVO$_3$ due to orbital broken symmetry enhanced by octahedral rotation

Although a metallic behavior would be expected for LaVO$_3$ ($t_{2g}^2 e_g^0$) with its two electrons distributed in three $t_{2g}$ partners, it is experimentally found to be insulating at all temperatures and no metal-insulator phase transition has been reported in its bulk form [6].
In order to understand the mechanism for gapping in this group, we initially break orbital occupancies by nudging two electrons in two out of three $t_{2g}$ orbitals on the V$^{3+}$ site in the assumed high symmetry cubic phase (*e.g.* (1,1,0) $t_{2g}$ occupancies). Following variational self-consistency, LaVO$_3$ reaches a lower total energy (ΔE=-321 meV/f.u *wrt* (2/3,2/3,2/3) degenerate cubic phase) with an insulating electronic structure having a band gap located between two occupied and one unoccupied formerly $t_{2g}$ levels (Fig.5.a). From this imposed orbital broken symmetry, we conclude that the electronic structure is unstable in the high symmetry cubic phase of rare-earth vanadates and that these materials should therefore have a secondary contribution to the band gap due to a Jahn-Teller distortion.

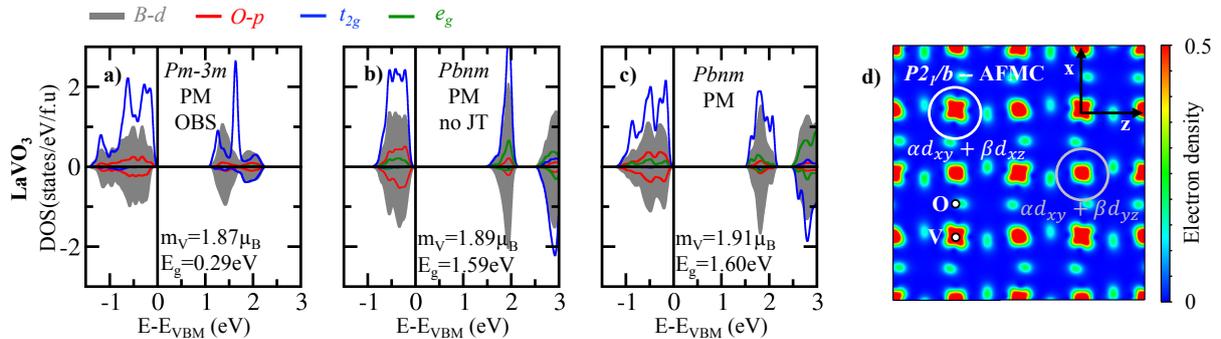

*Figure 5*: **a-c)** Averaged projected density of states on V d levels (grey) and O p levels (red) in LaVO$_3$ in the PM phase with different symmetries, lattice distortions or orbital broken symmetries (OBS). Projected density of states on $t_{2g}$ (blue) and $e_g$ (green) levels for a specific V cation within the supercell are also reported. d) Partial charge density plot in the (xz)-plane of the last 2 occupied bands in the AFM-C phase of LaVO$_3$.

This can be verified by allowing LaVO$_3$ to lower its energy within the PM spin order, by letting it develop O$_6$ rotations and antipolar motions, resulting in an orthorhombic symmetry.

This structure develops a sizable in-phase Jahn-Teller motion similar to that displayed by LaMnO$_3$, albeit smaller (TableSI5 in Supplementary S7), and is insulating with a large band gap of 1.60 eV located between $t_{2g}$ levels (Fig.5.c). Again, the JT distortion is but secondary for the gap opening, as evidenced by the fact that a *Pbnm distorted structure in which we artificially eliminated the JT distortion* already exhibits similar characteristics in terms of magnetic moments and band gap value (Fig.5.b). Rotations driven by pure steric effects, showing energy gain around 700 meV/f.u *wrt* the degenerate cubic phase (compared with the 321 meV/f.u produced by the intrinsic electronic instability), are again sufficient to alleviate orbital degeneracies.

At low temperatures LaVO$_3$ transforms to an AFM-C insulator. This LT phase develops an alternative and sizable Jahn-Teller motion (Table.SI5), for the octahedral distortion is in antiphase along the *c* axis while it is in-phase for the JT motion appearing in *Pbnm* phases (Fig. SI1 in Supplementary S1). This alternative JT motion lowers the symmetry from *Pbnm* to *P2$_1$/b* and produces a specific orbital pattern, different from that appearing in *Pbnm* phases (Fig.5.d and Fig.SI5 in Supplementary S9). The extra *d* electron *wrt* YTiO$_3$ is localized in a rock-salt pattern of an orbital corresponding of either a $\alpha d_{xy} + \beta d_{xz}$ or a $\alpha d_{xy} + \beta d_{yz}$ ($\alpha^2 + \beta^2$=1) combination. The orbital pattern favors an AFM-C order through the Kugel-Khomskii mechanism [40] without significantly changing the band gap and magnetic moments (Fig.2).

We see that due to pure steric effects, the O$_6$ rotations are sufficiently large to produce a new basis of orbitals, rendering a localized state in the *Pbnm* symmetry. The JT motion appearing in the orthorhombic PM phase is not electronically driven, in a similar fashion to LaMnO$_3$, while that of the LT phase is reminiscent of the native electronic instability of these materials [22] and of the "orbital-order" phase transition reported experimentally [6]. Finally, the observation of insulating phase in the orthorhombic symmetry without Jahn-Teller motion is again compatible with DMFT simulations of Pavarini *et al* on YVO$_3$ and LaVO$_3$ [18].

### d. Mechanism IV: Compounds with unstable single local electronic occupation pattern disproportionating into a double local environment: CaFeO$_3$ and YNiO$_3$

Although CaFeO$_3$ and YNiO$_3$ develop an electronic degeneracy similar to that of LaMnO$_3$ with a single $e_g$ electron, we note that CaFeO$_3$ and YNiO$_3$ are metallic within the orthorhombic

*Pbnm* PM phase, and become insulating in a lower symmetry space group – the *P2$_1$/n* monoclinic structure. We present here details for YNiO$_3$, the very same conclusions are drawn for CaFeO$_3$ in Supplementary S10.

We start from an ideal *Pm-3m* cubic phase and artificially offer breaking of the degeneracy of the Ni$^{3+}$ $t_{2g}^6 e_g^1$ levels by forcing a specific $e_g$ partner occupancy (*e.g.* (1,0) instead of (0.5,0.5)). However, the imposed orbital broken symmetry does not survive the variational self-consistency and the $e_g$ electron spreads equally on the two orbitals. We offer an additional symmetry breaking route by breaking of the degeneracy of the Ni$^{3+}$ levels *via* forcing $e_g$ levels occupancies on two different Ni sites to be (1,1) and (0,0) respectively. This yields a rock-salt pattern of Ni sites with half-filled and empty $e_g$ levels, respectively. Following variational self-consistency, YNiO$_3$ is trapped in such a state that proves to be of lower energy by 15 meV/f.u *wrt* the degenerate cubic phase, still with degenerate $e_g$ levels on each Ni site and no gap (Fig.6.a). In this OBS, we detect slightly different electronic structures between neighboring transition metal sites (labelled Ni$_L$ and Ni$_S$) yielding different $e_g$ level occupancies and magnetic moments (Fig.6.c). We determine from this probing of the linear response of the system to orbital nudging that the electronic structure is latently unstable and prone therefore to yield distortions in order to produce two types of B environments, one with half-filled and one with empty $e_g$ states.

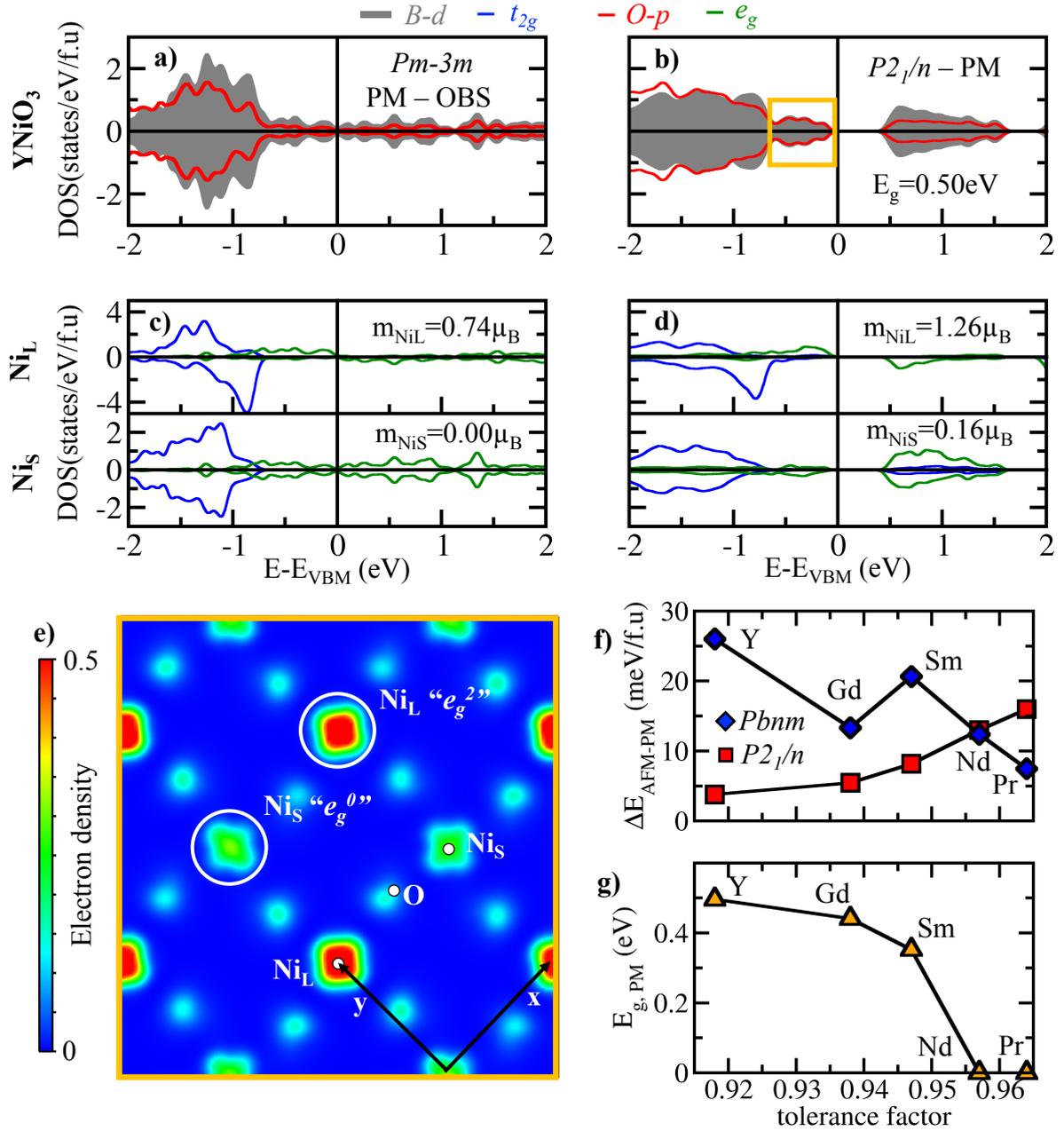

*Figure 6:* (a,b) Averaged projected density of states on Ni *d* levels (grey) and O *p* levels (red) in YNiO$_3$ in the cubic (a) and monoclinic (b) PM phase. (c,d) Projected density of states on $t_{2g}$ (blue) and $e_g$ (green) levels for a couple of Ni$_L$ (upper panel) and Ni$_S$ (lower panel) cations within the supercells. e) Partial charge density maps in the (*ab*)-plane of levels located near the Fermi level reported in b. f) Energy difference (in meV/f.u) between the AFM ground state and PM solutions using the orthorhombic (blue diamond) and monoclinic (red squares) symmetries as a function of the tolerance factor. G) Band gap E$_g$ (in eV) associated with the lowest PM phase as a function of the tolerance factor.

Building on our determination of the role of electronic instability, we can now follow-up performing the structural relaxation starting from the cubic phase with OBS. We find that YNiO$_3$ relaxes to a *P2$_1$/n* monoclinic phase (ΔE=-944 meV/f.u energy lowering *wrt* cubic phase) that is insulating with a gap of 0.50 eV (Fig.6.b). This phase is described by the usual O$_6$ rotations but it also develops a striking feature: there is a bond disproportionation of two O$_6$ groups, producing a rock-salt pattern of collapsed and extended octahedral (for which Ni cations sitting at their center are labeled Ni$_S$ and Ni$_L$, respectively) due to a sizable breathing mode (Table.SI5 in Supplementary S7), resulting in a double local environment (DLE) for B cations, each having very different B-O bonds. This disproportionation was described in detail by Park, Marianetti and Millis in terms of an 'intriguing correlated effect', but as seen here, it is entirely predictable by static mean field DFT [41]. The DLE results in different electronic structures for Ni$_L$ and Ni$_S$ ions, characterized by two magnetic moments, the former larger than 1 and the latter approaching 0 (Fig.6.b and d). Consistently with the breathing mode and the two magnetic moments, we find that the electrons prefer to occupy both *e$_g$* levels of Ni$_L$ cations as inferred by the partial charge density map of Figs.6.e. It yields a so-called "charge ordered" picture in which Ni$_S$ and Ni$_L$ cations have empty and half-filled *e$_g$* levels, respectively, getting rid-off the electronic degeneracy between neighboring Ni$^{3+}$ sites [42]. Therefore, consistent with the spontaneous electronic instability identified in the cubic phase, YNiO$_3$ reaches insulation through charge disproportionation effects transforming the 3+ unstable formal oxidation state (FOS) of Ni ion, to its more stable 2+/4+ stable FOS. Remarkably, although charge disproportionation occurs in the "3d + ligand" unit cell, the physical charge density $\rho(\vec{r})$ around Ni$_S$ and Ni$_L$ cations are nearly indistinguishable, owing to a charge self-regulation mechanism [43–45], whereby the ligands resupply the 3*d* site with charge lost. This result has sometimes been interpreted in terms of a 'ligand-hole' language [23,41,46].

We can finally check the relative role of rotations *wrt* the disproportionation effects (we recall that, alone, the latter does not open a gap in YNiO$_3$). To that end we have performed additional calculations for other RNiO$_3$ members (R=Gd, Sm, Nd and Pr) in which O$_6$ rotations progressively decrease (detailed results are presented in Supplementary S11). Surprisingly, we find that only R=Gd and R=Sm are relaxing to a monoclinic phase with a DLE, while R=Nd and Pr are more stable within the orthorhombic symmetry and a Single Local Environment (SLE) (Fig.6.f). Moreover, only R=Y, Sm and Gd become insulating while the other two compounds remains metallic (Fig.6.g). *We conclude that the O$_6$ rotation amplitudes are controlling the ability to disproportionate to an insulating PM phase in nickelates.*

However, we find that all the considered nickelates exhibit an AFM-S order, based on ↑↑↓↓ spin chains of Ni cations in the (*ab*)-plane with different stackings along the *c* axis [47], in their ground state. It thus produces a monoclinic *P2$_1$/n* symmetry that is insulating. We observe that the AFM-S phase shares all the key features of the monoclinic *P2$_1$/n* PM phase except the fact that the magnetic moment associated with Ni$_S$ cations becomes exactly null. The complex AFM-S order, which is compatible with the symmetry of the breathing mode, is therefore crucial to open the band gap *via* forcing disproportionation effects when materials develop small O$_6$ rotations.

Compounds with unstable electronic configuration in a SLE structure therefore reach insulation through several sequential factors: (i) they natively possess an electronic instability yielding disproportionation effects already in the high symmetry cubic phase to get rid-off the B cation unstable electronic configuration; (ii) the disproportionation is strongly linked to the octahedral rotations amplitude, appearing first due to steric effects (ΔE=-921 meV/f.u *wrt* the degenerate cubic phase in YNiO$_3$), and (iii) antiferromagnetic interactions force the disproportionation when O$_6$ rotations are weak. Our results reconcile the experimental phase diagram of rare-earth nickelates, and most notably, the PM-metal to PM-insulator or PM-metal to AFM-insulator as a function of the rare-earth ionic radius [12]. Moreover, we unify the different models proposed to explain the metal-insulator phase transition (MIT) with (i) the existence of an electronic instability in rare-earth nickelates (DMFT [48]) and (ii) a structurally triggered Peierls MIT in RNiO$_3$ and AFeO$_3$ (A=Ca, Sr) compounds, although the DFT calculations were restricted to simple spin-ordered states [25,26].

## *V.* Discussion

Our DFT calculations of both LT magnetically ordered and HT spin disordered PM phases reveal the minimal theoretical ingredients required to explain the trends in metal-nonmetal behavior in oxide perovskites, and the associated trends in forms of magnetism and structural selectivity. This includes spin polarization, magnetic interactions and lower energy phase search through the polymorphous representation—allowing large enough (super) cells so that various modalities of structural and electronic symmetry breaking can exercise their ability to lower the total energy. This leads one to identify four generic mechanisms opening a band gap in oxide perovskites. Two mechanisms are related to purely structural symmetry breaking such as natural octahedral symmetry, and its associated symmetry lowering distortions such as octahedral tilting and rotations due to steric effects. The two other mechanisms are related to intrinsic electronic instabilities of transition metals, originating either from spontaneous orbital symmetry breaking, *i.e* Jahn-Teller effect or unstable formal oxidation state of transition metals (bond disproportionation effect), both manifested by dedicated structural distortions.

***Consequences:*** Previous conclusions that DFT fails in describing gaping trends in Mott insulators were premature, and often based on a naïve application of DFT without properly exploring channels of energy- lowering symmetry- breaking by the modalities indicated in Section I, rather than on the failure of the description of interelectronic interactions underlying the density functional theory itself. Indeed, the sequential steps of offering to the system symmetry breaking modes appear to explain gaping, the nature of the magnetic order, and both the space group symmetries as well as sublattice distortions (Jahn-Teller and octahedral rotations) in both LT and HT phases of all $ABO_3$ perovskites with B=3$d$. This approach does not require explicit dynamical correlations, or the Mott picture of electron localization on atomic sites with the ensuing formation of upper and lower Hubbard bands (Fig.1). To further assess the role of interelectronic correlation in the sense of the Hubbard Hamiltonian on the band gap opening process, it is highly significant that calculations using SCAN meta-GGA functional without any kind of "U, systematically produce sizable band gaps for all compounds found insulating by experiments (Supplementary S3 details SCAN density of states, band gaps and magnetic moments). This shows that the basic interactions crystal field splitting, lattice distortions and spin-polarizations are sufficient to produce insulation in these compounds, and the celebrated Mott-Hubbard *mechanism* for gap opening (Fig.1.a) may not apply so generally to perovskite oxides. Thus, although these oxides are certainly complicated, they are not obviously strongly correlated materials.

**Implications:** We have identified that oxygen cage rotations – and lattice distortions in general – are a common underlying mechanism behind the gapping of many of these oxides and that their magnitude clearly determine whether or not the electrons will form a localized state. Many pathways to control the oxygen rotations are available and include pressure effects (homogeneous to the chemical pressure induced by A site cations) and interfacial effects through octahedral connectivity between different materials [49]. Beyond these "conventional" structural features, studying how defects and/or doping effects alter the properties of the compounds is another fundamental aspect. Modelling such entangled physics require techniques treating all degrees of freedom, and up to date only DFT can do it within a reasonable accuracy and a low computational cost.

**Acknowledgments**: This work of JV has been supported by the European Research Council (ERC) Consolidator grant MINT under the Contract No. 615759. Calculations took advantages of the Occigen machines through the DARI project EPOC A0020910084 and of the DECI resource FIONN in Ireland at ICHEC through the PRACE project FiPSCO. The work of AZ was supported by Department of Energy, Office of Science, Basic Energy Science, MSE division under Grant No. DE-FG02-13ER46959 to CU Boulder. We acknowledge discussions with G. Trimarchi and technical support from A. Ralph at ICHEC supercomputers.